\documentclass[10pt,letterpaper,twocolumn]{article} 

\usepackage[usenames,dvipsnames]{color}
\usepackage{ol2}
\usepackage{amsmath,array,amssymb}
\usepackage{graphicx}
\usepackage{hyperref}
\usepackage{textcomp}
\usepackage{braket}
\usepackage{multirow}
\usepackage{indentfirst}
\usepackage[FIGTOPCAP,raggedright,nooneline,bf,footnotesize]{subfigure}
\usepackage{paralist}
\usepackage{overpic}
\renewcommand{\eqref}[1]{Eq.~(\ref{#1})}

\newcommand{\figref}[1]{Fig.~\ref{#1}}

\newcommand{\kp}[1]{{\pmb{k}}_{#1 \perp}}
\newcommand{\sinc}{\text{sinc}}
\newcommand{\tpf}{\tau_p^\text{FWHM}}
\newcommand{\tp}{\tau_p}

\begin{document}
\twocolumn[ 
\title{Spectral correlation in down-converted photon pairs at telecom wavelength }
\author{ Andrzej Gajewski$^{1}$, Piotr Kolenderski$^{1,*}$}
\address{
$^1$ Faculty of Physics, Astronomy and Informatics, Nicolaus Copernicus University, Grudziadzka 5, 87-100 Torun, Poland\\
$^*$Corresponding author: kolenderski@fizyka.umk.pl
}

\begin{abstract}
Sources of photon pairs based on the spontaneous parametric down conversion process are commonly used for long distance quantum communication. The key feature for improving the range of transmission is engineering their spectral properties. Following two experimental papers {[}Opt. Lett., 38, 697 (2013){]} and {[}Opt. Lett., 39, 1481 (2014){]} we analytically and numerically analyze the characteristics of a source. It is based on a $\beta$ barium borate (BBO) crystal cut for type II phase matching at the degenerated frequencies $755$ nm $\rightarrow 1550$ nm + $1550$ nm.  Our analysis shows a way for full control of spectral correlation within a fiber-coupled photon pair simultaneously with optimal brightness.
\end{abstract}

\ocis{190.4410,300.6190,270.4180,270.5565}
] 

\maketitle

Sources of polarization entangled photon pairs based on the process of spontaneous parametric down conversion (SPDC) \cite{Uren2007,Hendrych2007,Valencia2007,Osorio2008,Evans2010,Eckstein2011,Gerrits2011, Horn2013} are now commonly used in various experiments such as those testing fundamentals quantum mechanics,  for quantum communication protocols or quantum  information processing. Each application defines a set of specific requirements for the characteristics of photons for the best performance. For example, for a heralded qubit encoded in a polarization state any correlation in any other degree of freedom, within a photon pair, will reduce the purity of the resulting state. Therefore there is a need to engineer characteristics of a source to produce uncorrelated pairs \cite{URen2005,Kolenderski2009,Florez2015}. In turn, the polarization entanglement is a basic block of quantum communication \cite{Gisin2002,Gisin2007} and quantum metrology \cite{Dayan2005,Dayan2007}.  The photons are transmitted through optical systems and single-mode fibers (SMF). 
Optical elements are dispersive and cause unwanted modification of the photons' temporal characteristics. The deteriorating dispersion effects can be reduced by using photon pairs that are spectrally narrow and feature appropriate spectral correlations \cite{Lutzthes2013,Lutz2013,Lutz2014}.

The spectra of SMF coupled photon pairs generated in SPDC process can be negatively \cite{Wasilewski2006,Jin2014}, positively \cite{Kim2002,Kuzucu2008,Lutzthes2013,Lutz2013,Lutz2014,Shimizu2009} or not correlated  at all \cite{Mosley2008,Jin2013,Bruno2014}. The effect of negative correlation originates from the energy conservation relation and is a consequence of pumping by a continuous wave (CW) laser. Uncorrelated spectra can be achieved  in a nonlinear crystal by using a pulsed laser pump with careful setup parameter choice or by using narrow spectral filters. However, for positive correlation, an additional requirement of group velocity matching needs to be fulfilled \cite{Kim2002}. 

\begin{figure}[h]
	\centering
	\includegraphics[width=0.95\linewidth]{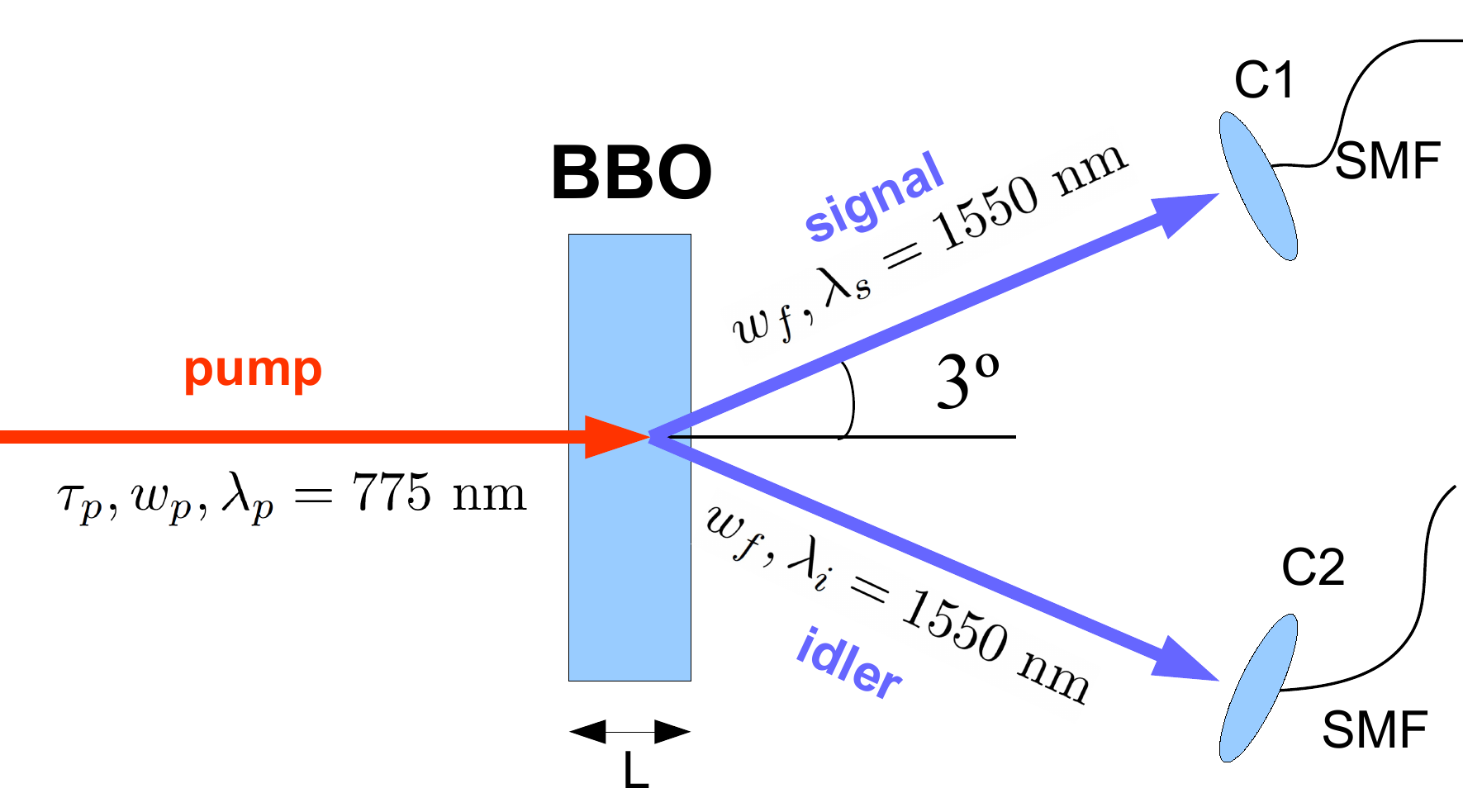}
	\caption{A single photon source is based on a $\beta$-barium borate (BBO) nonlinear crystal of thickness $L$ cut for type II phase matching at frequencies  $755$ nm $\rightarrow 1550$ nm + $1550$ nm and half cone opening angle $3$ deg. A pump pulse of $\tp$ duration is centered around $775$ nm and its transverse spatial mode is assumed to be a Gaussian function with a characteristic diameter  $2 w_p$. The photon pairs are coupled into single mode fibers (SMF), which together with  coupling stages (C1,C2) define a collected transverse spatial modes, which are assumed to be a Gaussian functions of diameters $2w_f$. } 
	\label{fig:setup}
\end{figure}

Here  we perform numerical analysis of the SPDC photon pair source experimentally analyzed in Refs \cite{Lutz2013, Lutz2014}, which is  schematically depicted in \figref{fig:setup}. We discuss trade-offs between the correlation parameter, pair production rate and SMF coupling efficiency, which are figures of merit allowing to estimate source's suitability for a given application. In the setup a femtosecond laser pulses centered at $775$ nm are used to pump a BBO crystal, which optical axis orientation is set at $29.68$ degree allowing type II phase matching at degenerate frequencies $1550$ nm. The two created photons propagating at an $3$ degree angle with respect to pump beam propagation direction are coupled into SMFs.

Firstly, the goal is to derive a simplified compact analytical formula for characteristics of our source. Following the mathematical model presented in Ref.~\cite{Kolenderski2009,Kolenderski2009a} we use an effective phase matching function (EPMF) $\Theta (\omega_s,\omega_i)$ in order to fully describe joint effect of the pump beam spatial profile  $A_p^{spat}(\kp{})$, the crystal's phase matching  $\sinc (\Delta k_z L /2)$  and the collected modes' profiles $u_s(\kp{s})$, $u_i(\kp{i})$. It is defined as the following overlap: $\Theta (\omega_s,\omega_i)=\int d\kp{s}d\kp{i} u_s(\kp{s}) u_i(\kp{i}) A_p^{spat}(\kp{s}+\kp{i}) \sinc (\Delta k_z L /2)$. Here $\Delta k_z=k_{pz}-k_{sz}-k_{iz}$ stands for a phase mismatch of pump and resulting photons. Then a state of a photon pair propagating in single mode fibers can be conveniently expressed as a product of a pump pulse spectral mode $A_p(\omega)=\sqrt{\tp}\exp\left(-\tp^2(\omega-2\omega_0)^2/2\right)$ and EPMF: $	\Psi (\omega_s,\omega_i) = A_p (\omega_s+\omega_i) \Theta (\omega_s,\omega_i)$. Within this definition the pulse duration $\tau_p$ is related to value at full width half maximum (FWHM) by $\tp=\tpf/\sqrt{8 \log(2)}$.

In order to get a simplified expression for the biphoton wave function $\Psi (\omega_s,\omega_i)$ we approximate pump and SMF's modes using Gaussian functions $A_p^{spat}(\kp{})=w_p\exp\left(-w_p^2\kp{}^2\right)$,  $u_{\mu}(\kp{s})=w_f\exp\left(-w_f^2\kp{\mu}^2\right), \mu=s,i$, respectively. The characteristic transverse mode diameters $2 w_p$ and $2 w_f$ are  measured at $13.5 \%$ of intensity maximum.  
Next, the phase mismatch can be expanded up to first order in Tylor series in angular and spatial frequencies \cite{Grice1997,Kolenderski2009} : $\Delta k_z \approx \beta_s (\omega_s-\omega_0) +\beta_i (\omega_i-\omega_0) + d_{sx} (k_{sx}-k_{sx0}) + d_{ix} (k_{ix}-k_{ix0}) $. For our particular setting the magnitudes of respective coefficients are numerically equal within a good approximation therefore we introduce the following simplifying assumption: $d_{sx} = - d_{ix}=d_x $
and $\beta_s=-\beta_i=\beta$.
 This together with the approximation for the sinc function  \cite{Kolenderski2009} $\sinc(x)\approx \exp(-x^2/5)$ allows us to derive a simplified analytical formula for EPMF and consequently for the biphoton, which is again in a Gaussian form: 
\begin{equation}
\Psi(\omega_s,\omega_i) = \mathcal{N} \exp\left(-\frac{(\omega_s-\omega_i)^2}{2\sigma_{+}^2}-\frac{\tp^2}{2}(\omega_s+\omega_i-2 \omega_0)^2\right)
\label{eq:psi}
\end{equation}
with a characteristic width given by:
\begin{equation}
	\sigma_{+}^2 =  \frac{1}{{\beta ^2}} \left(\frac{2 d_x^2}{w_f^2}+\frac{5}{L^2}\right),
\label{eq:width}
\end{equation}
where $\mathcal{N}$ stands for a scaling factor. Note that the wave function within this approximation is real. In the following part we juxtapose the approximate analytical model and precise numerical simulation results in order to discuss conditions for optimal performance of a source.

When a nonlinear crystal is pumped with a CW laser, the negative correlation of signal and idler photons' frequencies originates in energy conservation relation. For pulsed pumping, the setup parameters can facilitate not only negative but also positive or no correlations. This, seemingly,  energy conservation relation breaking, is a result of a pump pulse spectral width and characteristics of down conversion process.

\begin{figure}[h]
	\centering
	\begin{tabular}{c}
    \subfigure[$\Theta (\omega_s,\omega_i)$, $L=1$ mm]{\includegraphics[width=0.48\linewidth]{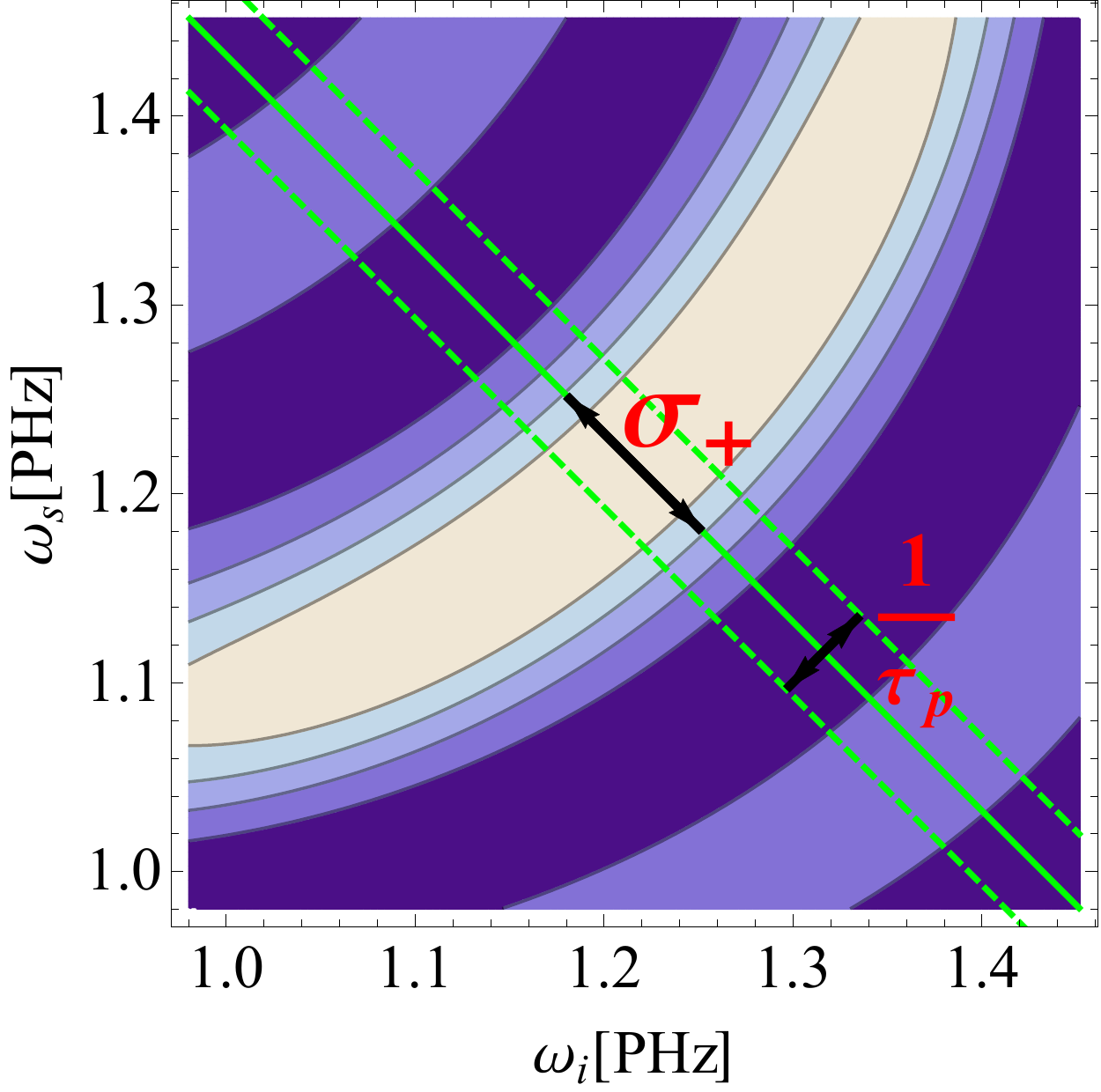}}
    \subfigure[$\Psi (\omega_s,\omega_i)$, $L=1$ mm]{\includegraphics[width=0.48\linewidth]{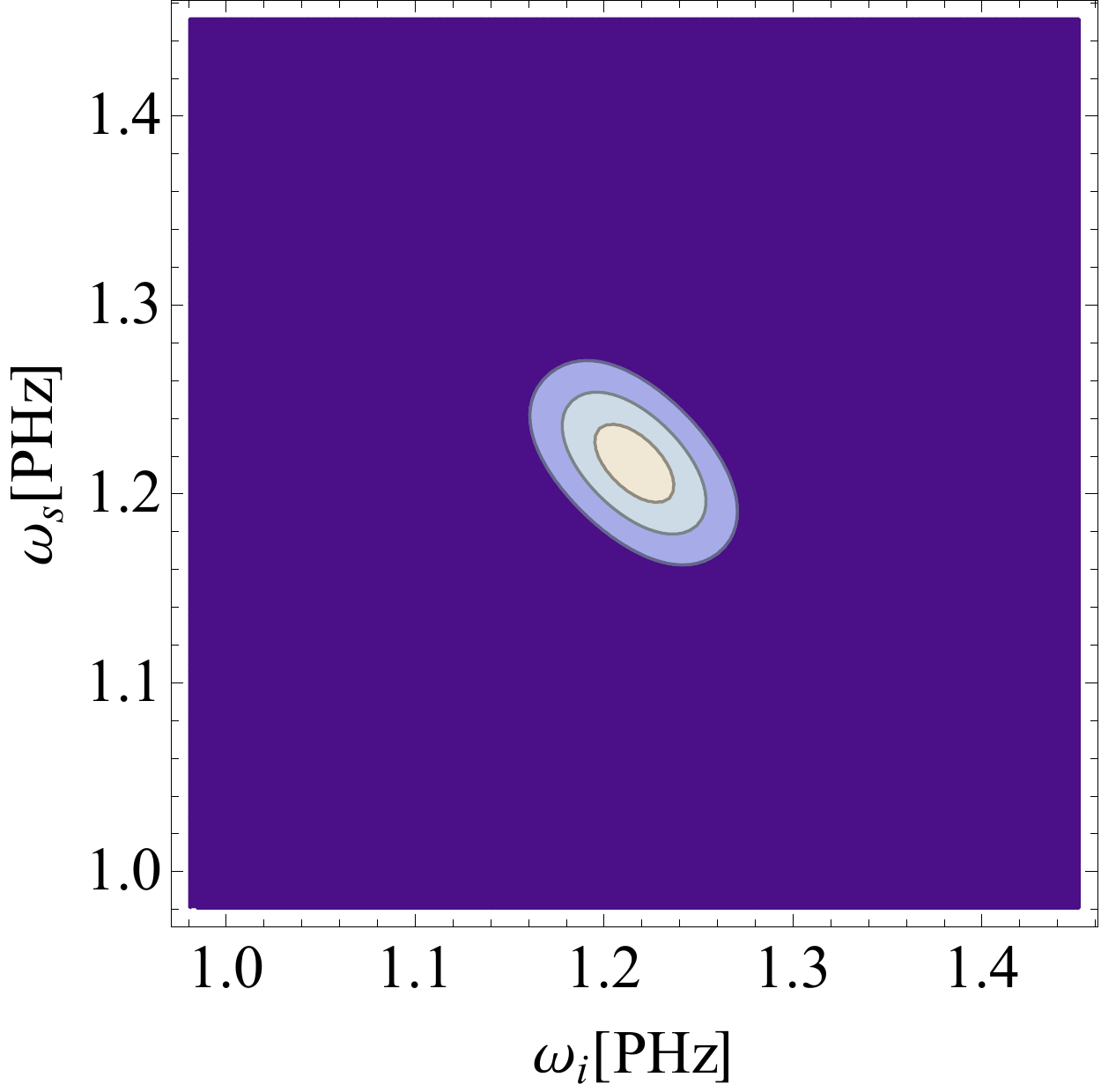}}\\
    \subfigure[$\Theta (\omega_s,\omega_i)$, $L=7.5$ mm]{\includegraphics[width=0.48\linewidth]{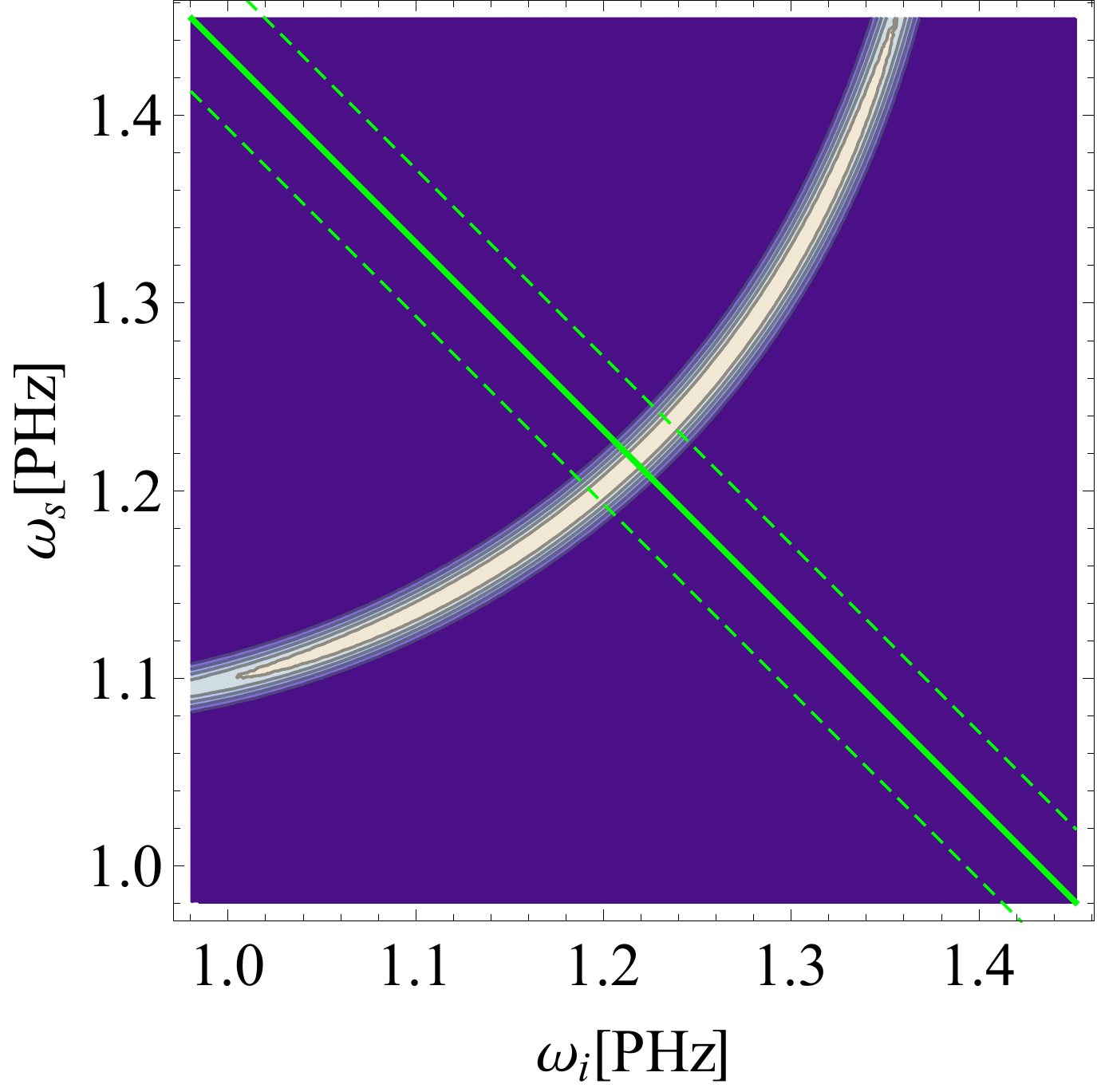}}
    \subfigure[$\Psi (\omega_s,\omega_i)$, $L=7.5$ mm]{\includegraphics[width=0.48\linewidth]{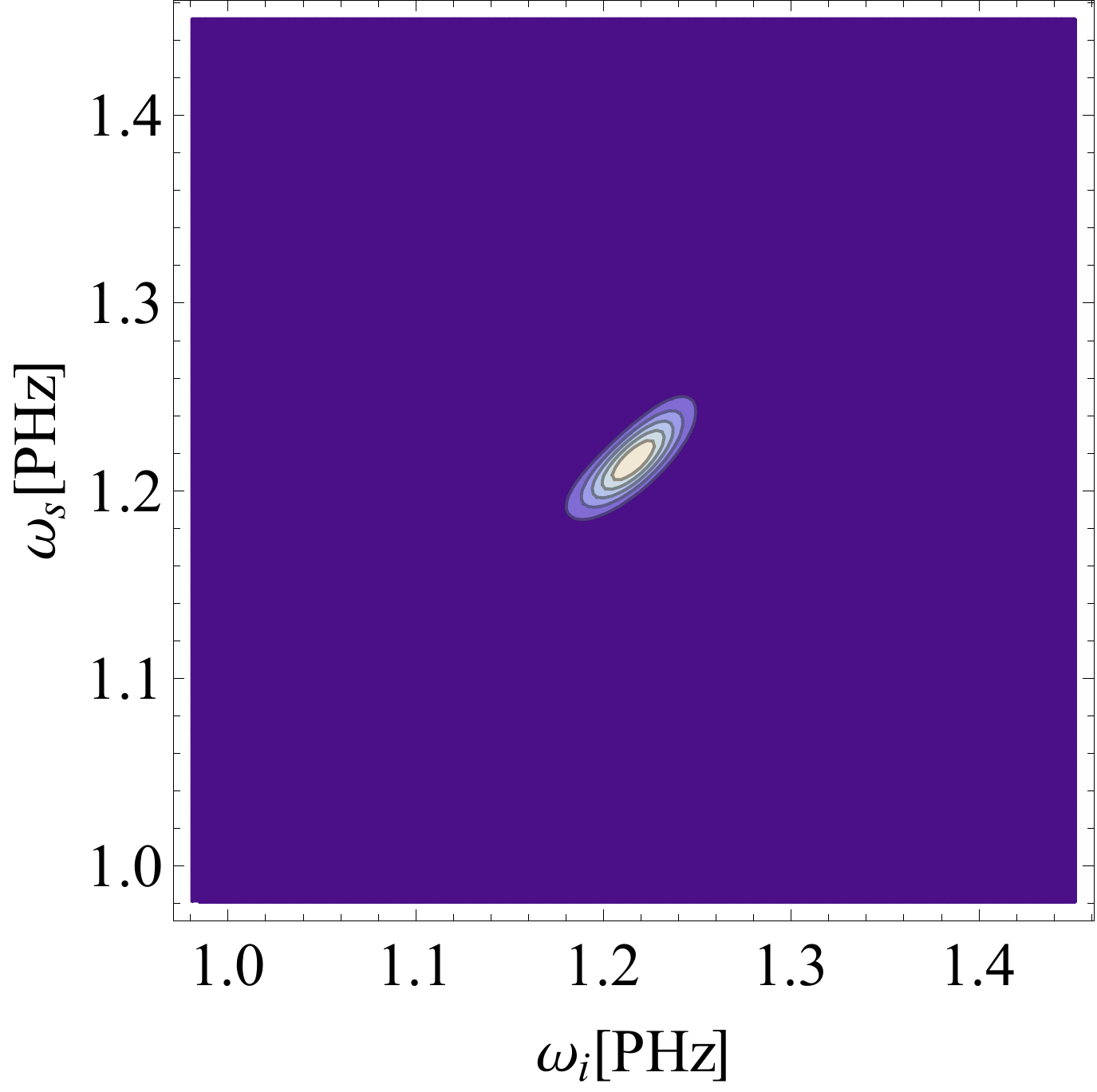}}
	\end{tabular}
	\caption{ (a,c) The effective phase matching function $\Theta (\omega_s,\omega_i)$ and (b,d) the biphoton wavefunction $\Psi (\omega_s,\omega_i)$ for two exemplary crystal lengths (a,b) $L=1$ mm and (c,d) $L=7.5$ mm are depicted with contours. The functions are real. Green solid (dashed) lines represent pump pulse central frequency (spectral width). The simulations were performed for: $\tpf = 50$ fs, $w_p= 100$ $\mu$ m, $w_f=440$ $\mu$m. 
	}
	\label{fig:EPMF}
\end{figure}

In \figref{fig:EPMF} we present how positive and negative correlation can be achieved. Orientation of the biphoton wave function depends on phase matching condition and pump spectral amplitude. For photons with the same energy, EPMF is a function which maximum is approximately oriented in antidiagonal direction in frequency space as depicted in \figref{fig:EPMF}(a). A pump's orientation is in the diagonal direction, which is depicted by the green  line and two green dashed lines representing pulse characteristic spectral width $1/\tau_p$. The biphoton wave function is a product of those two \cite{Grice1997} as seen in \figref{fig:EPMF}(b).

We quantify the spectral correlation using the Pearson coefficient, $r$, and taking the probability distribution based on the biphoton wave function  \eqref{eq:psi}: 
\begin{equation}
r= \frac{\langle(\omega_s-\omega_0) (\omega_i-\omega_0)\rangle}{\sqrt{\langle(\omega_s-\omega_0)^2\rangle\langle(\omega_i-\omega_0)^2\rangle}} = \frac{1-\tp^2 \sigma_+^2}{1+\tp^2 \sigma_+^2}.
\label{eq:r} 
\end{equation} 
We observe that a correlation is negative (positive) if the quantity $\sigma_+ \tp$ is $>1$ ($<1$). There is no correlation for $\sigma_+ \tp=1$. The type depends on the relative widths of EPMF, $\sigma_+$, and pulse spectral amplitude, $1/\tau_p$ . If the width of EPMF is larger (smaller) compared to pump's as in \figref{fig:EPMF}(a) (\figref{fig:EPMF}(c)) the resulting biphoton spectral state is negatively (positively) correlated as in \figref{fig:EPMF}(b) (\figref{fig:EPMF}(d)). If both widths are equal, photons feature no correlation at all.

\begin{figure}[h]
	\centering
		\begin{tabular}{c}
	    \subfigure[ $L=7.5$ mm, $\tp= 150$ fs]{\includegraphics[width=0.94\linewidth]{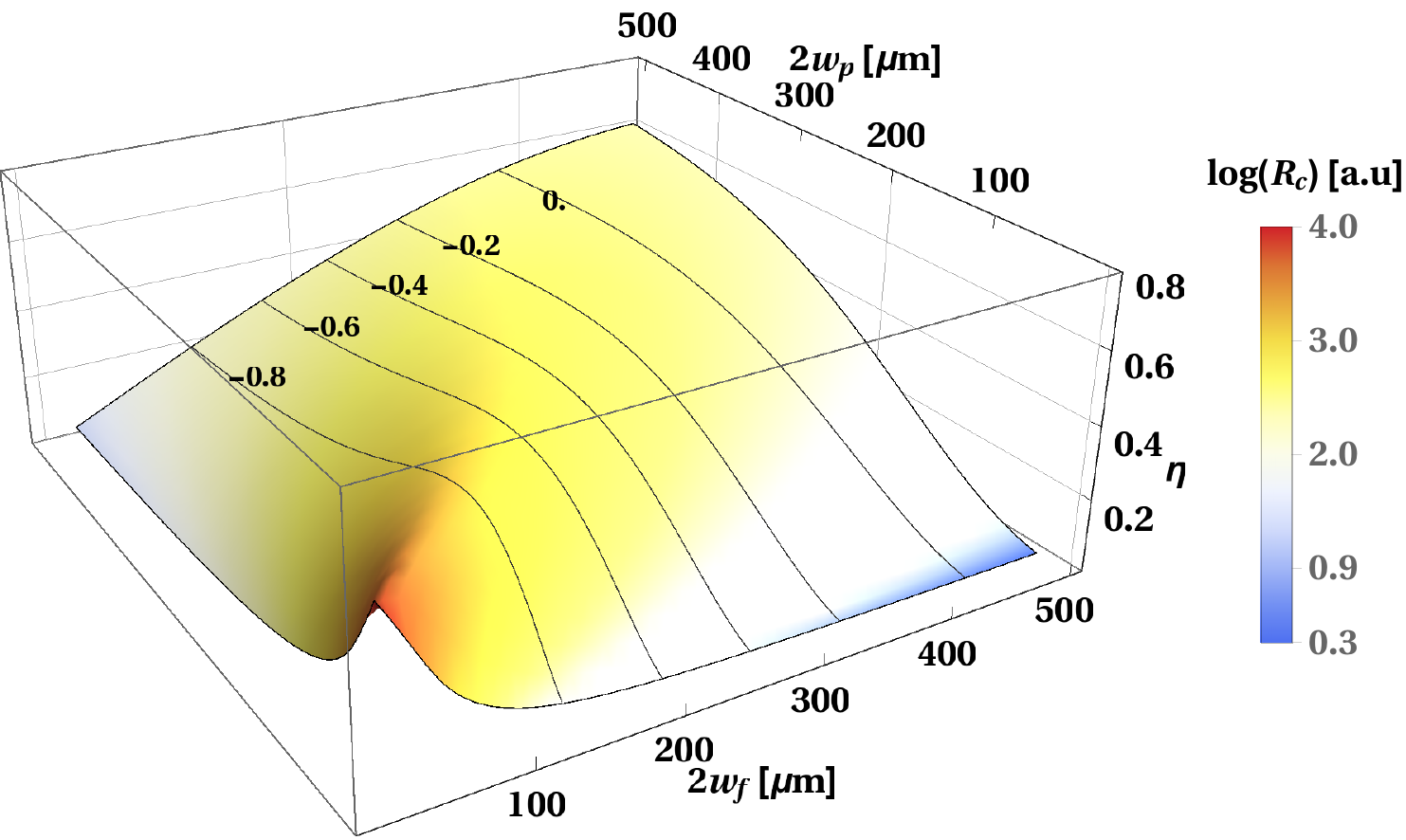}}\\
	    \subfigure[ $L=7.5$ mm, $\tp= 50$ fs]{\includegraphics[width=0.94\linewidth]{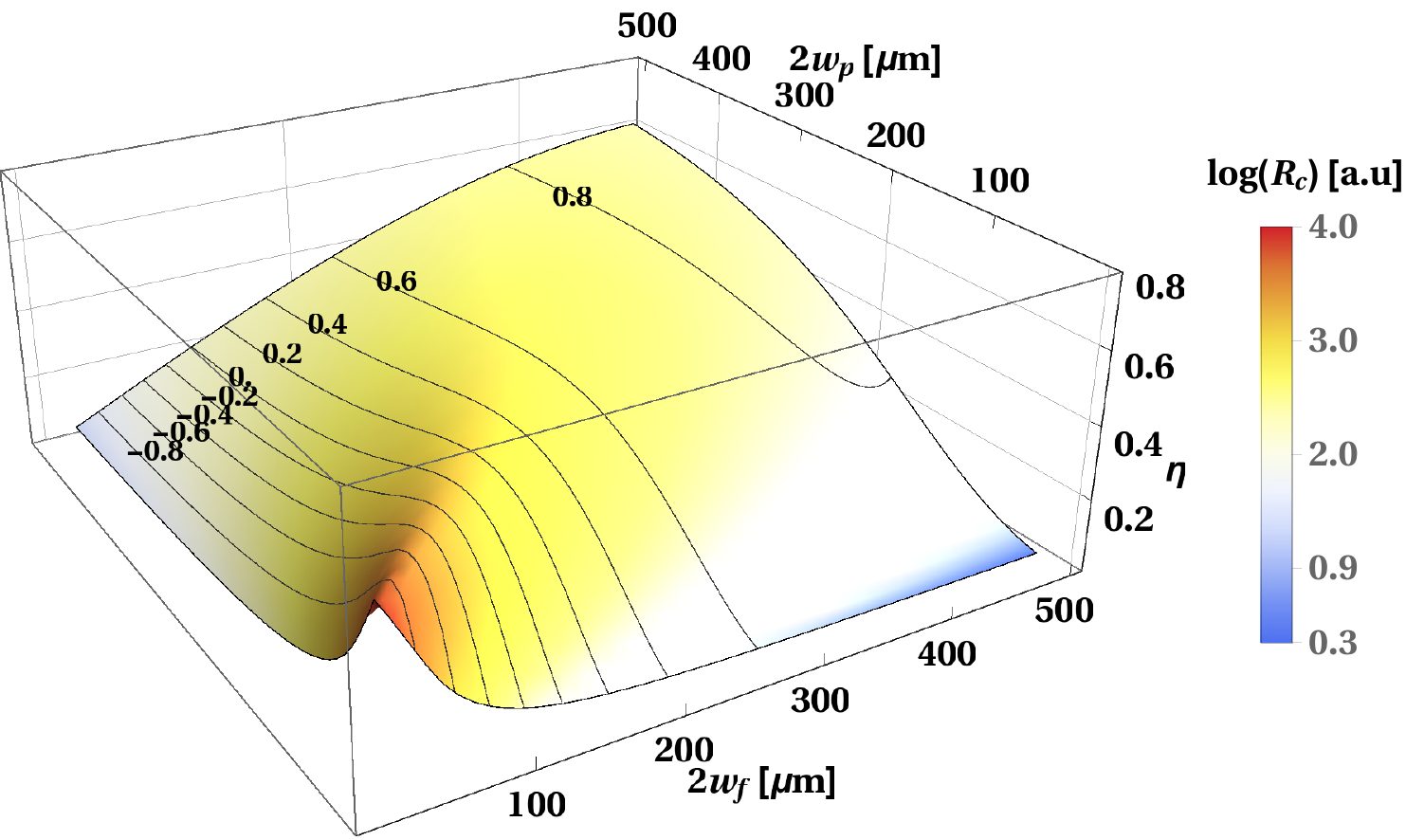}}\\
	    \subfigure[ $L=2.5$ mm, $\tp= 50$ fs]{\includegraphics[width=0.94\linewidth]{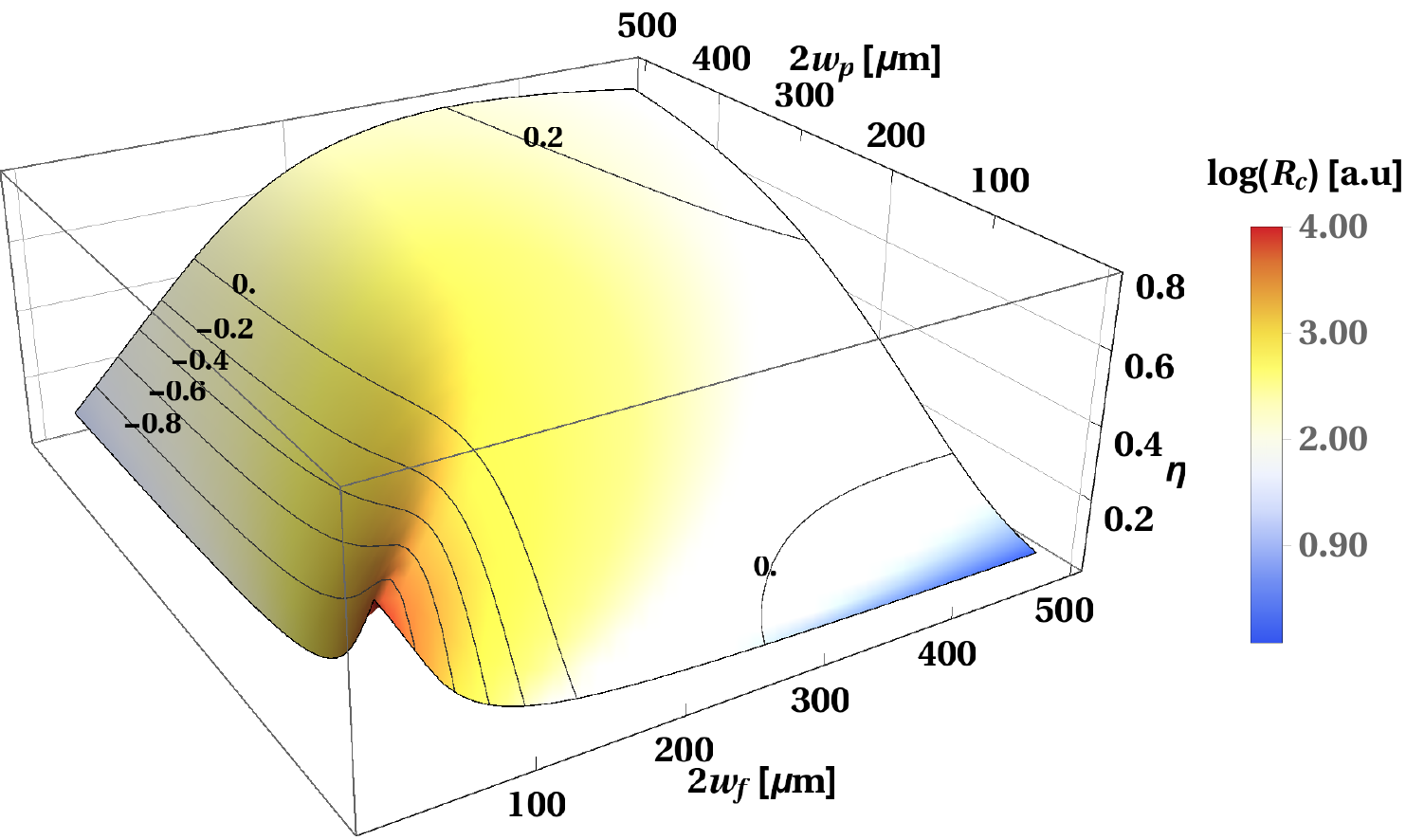}}\\	    
		\end{tabular}
	\caption{The SMF coupling efficiency $\eta$ (surface plot), SMF coupled pair production rate $R_c$ (color scaling) and correlation parameter $r$ (contours) as a function of pump $2w_p$ and collected $2w_f$ spatial mode diameters. 
	} 
	\label{fig:map}
\end{figure}

Our analytical model shows that correlation parameter $r$, introduced in \eqref{eq:r} when combined with the definition \eqref{eq:width}, is a function of the pump pulse duration $\tp$, the length of the crystal $L$ and collected mode width $w_f$ only. It does not depend on the pump spatial mode width $w_p$. Let us analyze the condition for spectrally uncorrelated photons $\sigma_+^2\tp^2=(\frac{2 d_x^2}{w_f^2}+\frac{5}{L^2})\tp^2=1$. It splits the space of setup parameters into regions of positive and negative correlation. The formula shows that the thinner the crystal or the narrower collected mode profile,  the shorter pulse is required for spectrally uncorrelated photons. This observation also agrees with our numerical simulation results shown  in \figref{fig:map}. We consider three exemplary settings involving two crystals thicknesses $L=7.5$ mm and $2.5$ mm and two pules durations $\tpf= 150$ fs, $50$ fs. The correlation parameter, $r$ is depicted using contours, which are almost parallel to $w_p$ axis for broad range of spatial mode profiles parameters. 

It is favorable for some applications when the SMF coupled photon pair exhibits significant positive spectral correlation. In order to achieve that, EPMF width $\sigma_+$ has to be small. One way of doing that is by using thick crystals as shown in \figref{fig:EPMF}(b,d). The analytical approximation  \eqref{eq:width} 
shows a dependence on collected mode width $2 w_f$ and crystal thickness $L$. In the limit of long crystals or large SMF collected mode profile, the EPMF's width $\sigma_+$ is always finite. Therefore there is a broad range of experimentally available parameters, where  $\sigma_+$ is optimal for positive spectral correlation. It can be seen  in \figref{fig:map}(a) that typical pulse duration of $150$ fs and thick crystal $7.5$mm allows for positive correlation as high as $r\approx 0.3$ are possible. We observe that using longer crystals $L>7.5$ mm with all other parameters fixed does not increase much the value of correlation parameter. On the other hand applying shorter pulse duration of $50$ fs allows one to get $r\approx 0.8$ ($0.2$) for $L=7.5$ mm ($2.5$mm) , see \figref{fig:map}(b,c). The use of shorter crystal is also beneficial in terms of SMF coupling efficiency, which we will analyze later. However it limits the magnitude of spectral correlation. 

As a side note, we point out that the spectra of signal and idler photons depend on a pules duration $\tau_p$ as well as EPMF's width $\sigma_+$. The widths, which are approximately equal to each other, can be derived based on \eqref{eq:psi} as $\sqrt{\frac{1}{2}\left(\frac{1}{\tp^2}+ \frac{2 d_x^2}{\beta^2 w_f^2}+\frac{5}{\beta^2 L^2}\right)}$. It is easy to see that conditions favorable for positive spectral correlation make the photons spectrally narrower for fixed pulse duration.


Now we move on to an analysis of spectral correlation in conjunction with other source's characteristics, which define its usefulnesses for certain applications. Those are: a SMF coupled pair production rate $R_c$ and a SMF coupling efficiency $\eta=R_c/\sqrt{R_s R_i}$, where $R_s$ and $R_i$ are signal and idler photon production rates, respectively. 
We numerically analyze  \cite{Kolenderski2009,Kolenderski2009a} how those source characteristics depend on spatial mode widths of the pump $w_p$ and collected mode $w_f$.  \figref{fig:map} shows our numerical simulation results: surface illustrates efficiency $\eta$ dependence,  black contours on the top of the efficiency  surface depict a correlation $r$ and color scaling is for a SMF pair production rate $\log (R_c)$ . 

The SMF coupled pair production rate $R_c$ depends on spatial modes of the pump $w_p$ and collected modes $w_f$. It increases very quickly with decreasing spatial mode widths, see red region in \figref{fig:map}(a,b,c) for small $w_f$ and $w_p$. However it is almost flat for a broad range of pump and collected mode widths (yellow region). We also observe it changes slowly with increasing crystal thickness $L$. This can be understood as a result of two competing effects: the longer the crystal, the higher the probability to convert a pump photon into a pair, but on the other hand, the stronger the phase matching requirement limiting the spectral range of photons.

Next, the SMF coupling efficiency, $\eta$, strongly depends on geometrical parameters, but exhibits weak dependence on spectral width of the pump $ {1}/{\tau_p}$. The results of our numerical simulations suggest that the highest coupling can be achieved when spatial modes of the pump and collected photons are both large and approximately the same. For $L = 7.5$ mm ($2.5$  mm)  and spatial modes diameters $2 w_p= 500 \mu$m  $2 w_f=500 \mu$m ($340 \mu$m) it reaches $\eta\approx0.6$ ($0.76$)  as can be seen  in \figref{fig:map}(a,c). We observe that the coupling efficiency, $\eta$, decreases when  crystal thickness $L$ gets bigger. This can be attributed to a transverse walk-off effect in birefringent type II crystal.

One can use \figref{fig:map} as a recipe for a source, which generates pairs of a given correlation parameter $r$. Let us assume that the crystal thickness and pulse duration are fixed. The  collected mode spatial profile is determined by a chosen correlation parameter. Next the optimal spatial pump mode can be chosen in order to achieve the best possible coupling efficiency or the SMF coupled pair production rate, $R_c$. Note also that $R_c$ does change slowly in a wide range of spatial mode widths $w_p$ and $w_f$, which is in favor of the described optimization method.

Summarizing, we discussed optimal settings for a  source capable of generating photon pairs in telecom range featuring any kind of spectral correlation. For a given crystal thickness it is possible to control the spectral correlation by adjusting the pump pulse  or collected modes characteristics. The correlation  was analyzed together with the SMF coupled pair production rate and coupling efficiency.  Our numerical simulations showed that there is an asymptotic limit for the length of the crystal yielding maximal pair production rate. We also presented a recipe allowing one to simply choose optimal source parameters for  a given correlation parameter and optimal both SMF coupled pair production rate and SMF coupling efficiency.
In particular, the analyzed source can generate pairs featuring positive correlation. In addition to that the spectra of each individual photons got narrower with increasing positive correlation value for a fixed pulse duration. These two facts make it useful in applications where dispersion effects matter such as long distance fiber based quantum communication. 

The authors acknowledge support by the National Laboratory FAMO in Torun, Poland, financial support by Foundation for Polish Science under Homing Plus no.~2013-7/9 program supported by European Union under PO IG project and by Polish Ministry of Science and Higher Education under Iuventus Plus grant no.~IP2014 020873.

\bibliographystyle{ol}

\end{document}